\documentclass[oldversion]{aa}

\usepackage{graphicx}
\usepackage{txfonts}

\begin{document}

\title{Signatures of star streams in the phase space distribution of nearby
 halo stars}
  
\author{C. Dettbarn\inst{1} 
          \and B. Fuchs\inst{1}
	  \and C. Flynn\inst{2,}\inst{3}
	  \and M. Williams\inst{3}}

\offprints{C. Dettbarn,
\email{dettbarn@ari.uni-heidelberg.de}}

\institute{Astronomisches Rechen-Institut am Zentrum f\"ur Astronomie der 
Universit\"at Heidelberg, M\"onchhofstr.~12 - 14, 69120 Heidelberg, Germany \\
\and
Tuorla Observatory, University of Turku,
 V\"ais\"al\"antie 20, FI-21500 Piikki\"o, Finland \\
\and
Mt.~Stromlo Observatory, Australian National University,
Cotter Rd., Weston Creek ACT 2611, Australia \\ }

\date{Received  2007}

\abstract{
We have analyzed the phase space distribution of a sample of about 900
non--kinematically selected low metallicity stars in the solar vicinity. The
stars primarily represent the thick disk and halo populations of the Milky Way.
We aim to identify overdensely populated regions in phase space, which we
interpret as signatures of star streams passing close to the Sun. The search was
conducted in a space constructed from the angular momenta and eccentricities of
the stellar orbits. Besides recovering all well known star streams in the thick
disk, we isolated four statistically significant phase space 
overdensities amongst halo stars. One of them is associated with a previously 
known halo star stream, but three of them are novel features, which we propose 
be also considered as genuine halo streams.}


\maketitle

\section{Introduction}

The recent large scale surveys of the stellar population of the Milky Way have 
revealed the presence of long coherent stellar streams. Most spectacular are 
the Sagittarius stream, which wraps around the entire Galaxy nearly 
perpendicular to the midplane, (Ibata et al.~2001, Mart\'inez-Delgado et
al.~2001, Majewski et al.~2003, Belokurov et al.~2006), and at low Galactic 
latitudes the Monoceros stream (Yanny et al.~2003, Ibata et al.~2003). 
These are interpreted as tidal debris associated with disrupted satellite 
galaxies, which were accreted by the Milky Way. Numerical simulations of 
such merger events have shown that such debris streams survive as coherent
structures over gigayears so that they are witnesses of the formation history 
of the Galaxy (Helmi 2004, Law, Johnston \& Majewski 2005, Pe\~narrubia, 
Mart\'inez-Delgado \& Rix 2005). 

Stellar streams have also been discovered as overdensities in the
phase space distribution of stars in the solar neighbourhood. Helmi et
al.~(1999) detected signs for a star stream in the halo population passing 
close to the Sun which was later confirmed by Chiba \& Beers (2000). 
Prominent streams or more correctly `moving groups' in the disk population of 
stars in the vicinity of the Sun are the Hercules, Pleiades--Hyades, and Sirius 
streams (Eggen 1996, Dehnen 1998). These moving groups are very probably not
related to tidal debris streams, but originate dynamically within the disk 
itself. 
Dehnen (2000) has demonstrated that stars in the Hercules group are presumably 
in outer Lindblad resonance with the inner bar of the Milky Way which leads to 
focussing of stellar orbits into certain parts of phase space. Quillen \& 
Minchev (2005) have suggested that the other moving groups are due to 
resonances with spiral density waves. All these 
features can be traced both in the thin and the thick disk populations. 
The Arcturus stream (Eggen 1996), on the other hand, is attributed to the
thick disk only and is probably of extragalactic origin (Navarro, Helmi \&
Freeman 2004). Arifyanto \& Fuchs (2006) have analyzed the kinematics of 
nearby subdwarfs and identified another stream in the thick disk which
resembles closely the Arcturus stream. The same feature has been detected 
independently by Helmi et al.~(2006) in the phase space distribution of nearby
F and G stars. The nature of this stream is at present unclear. Helmi et 
al.~(2006) argue in favour of an extragalactic origin, because they find that
its phase space structure resembles that of numerically simulated debris 
streams of disrupted satellite galaxies. 
\begin{figure*}
\centering
\caption{Slices through the phase space distribution of halo stars in the 
solar neighbourhood. The panels a, b, c, and d show the velocities of stars on
orbital planes inclined relative to the direction towards the North Galactic 
Pole by angles of $\nu$ = $0^\circ - 45^\circ$, $45^\circ - 90^\circ$,
$90^\circ - 135^\circ$, and $135^\circ - 180^\circ$, respectively (top left
to bottom right). 
$V_{\rm az}$ denotes the tangential velocity of the stars and 
$V_{\Delta{\rm E}}$ is a measure of the eccentricities of the stellar orbits 
(cf.~section 2). }
\label{fig1}
\end{figure*}
On the other hand, it could as well owe its
existence to similar dynamical effects which caused the Hercules group. If the
members of the Hercules group are at outer Lindblad resonance with the Galactic
bar (Dehnen 2000), the members of the new group would be at the outer 4:1 
ultraharmonic resonance with the bar (Fuchs, unpublished). These 
questions can only be settled, if the elemental abundance pattern of the 
members of the streams are known. They would indicate that either the stars 
belong to the field population of stars in the Milky Way or to the 
population of a disintegrated dwarf galaxy with different abundances 
(Williams \& Freeman, in preparation). 

In this study we investigate the phase space distribution of halo stars
in the solar neighbourhood, searching for signatures of streams in the Galactic
halo.

\section{Search Strategy for Star Streams in the Halo Population}

Our analysis is based on the sample of non--kinematically selected low 
metallicity stars of Beers et al.~(2000). We have drawn from that catalogue the 
space velocity data of all stars with metallicties [Fe/H] $< -1$, which leads 
to a sample of 897 stars. To the heliocentric space velocity vector $(U, V, W)$
of each star, with $U$ pointing towards the Galactic center, $V$ into the 
direction of Galactic rotation, and $W$ towards the North Galactic Pole,
respectively, the 
velocity vector of the motion of the Sun relative to the local standard of
rest $(U_\odot, V_\odot, W_\odot)$ = $(10.0, 5.3, 7.2)$ km/s was added 
(Dehnen \& Binney 1998). Next, the velocities are transformed 
from the reference frame of the local standard of rest to an inertial frame 
by adding to the $V$ velocity components the circular velocity of the local 
standard of rest around the Galactic center, for which we adopt the value 
of $V_{\rm LSR}=220$ km/s. In the following we assume a spherical
Galactic gravitational potential and neglect the flattening of the disk 
potential and any asphericity of the dark halo potential. Chiba \& Beers (2000)
have modelled the Galactic gravitational potential with aspherical St\"ackel
potentials. They showed that the distribution of halo stars in the phase space 
spanned by the isolating integrals of motion of the stellar orbits in such 
a potential can be closely mapped into the corresponding phase space 
constructed assuming a spherical potential, which justifies our procedure. 

In a spherical potential a star moves in a constant orbital plane. Most of the
stars which we have extracted from the Beers et al.~(2000) sample have 
distances less than 1 kpc from the Sun, and for the purpose of calculating
orbital parameters may be assumed to be at the Sun. 
The azimuthal velocity of a star is obviously given
by $V_{\rm az}=\sqrt{(V+V_{\rm LSR})^2+W^2}$ and the angular momentum is then 
$L=R_\odot\cdot V_{\rm az}$, where  $R_\odot$ denotes the galactocentric 
distance of the Sun for which we assume $R_\odot$ = 8 kpc. The inclination
angle of  the orbital plane relative to the direction towards the North 
Galactic Pole is given by $\nu=\arctan{((V+V_{\rm LSR})/W)}$. We restrict in 
our 
analysis the inclination angles to the range $0^\circ \leq \nu < 180^\circ$. 
Stars in an orbital plane with an inclination angle larger than $180^\circ$ 
are treated as retrograde stars in the orbital plane with an inclination angle 
$\nu-180^\circ$. 
\begin{figure*}
\centering
\caption{Slices through the wavelet transform of the phase space distribution 
of halo stars in the solar neighbourhood constructed analogous to the slices
in Fig.~1. The phase space densities are colour
coded at 10 \% levels according to a linear colour table ranging from red,
over yellow, green, blue to dark violet. The labels indicate halo star 
streams. }
\label{fig2}
\end{figure*}
The aim of our study is to identify star streams among the 
halo stars. Thus we are searching for groups of stars on essentially the same 
orbits which pass at present through the solar neigbourhood. For this purpose 
we have generalized the search technique of Arifyanto \& Fuchs (2006) and have
grouped the stars in our sample  according to the inclination angles of their
orbital planes. In each $\nu$--slice we have then searched for overdensities 
in a space spanned by angular momentum, which defines the mean guiding
center radius of a stellar orbit, and eccentricity, which
describes the radial excursion of the orbit from its mean guiding center radius.
Since all stars are at present close to the Sun, they all have the same
phase along their orbits. Using the theory of Galactic orbits by Dekker (1976)
Arifyanto \& Fuchs (2006) have shown that the search can be conducted in
practice in the space of $V_{\rm az}$ and $V_{\Delta{\rm E}}=
\sqrt{U^2+2\,(V_{\rm az}-V_{\rm LSR})^2}$.
$V_{\rm az}$ is directly related to the angular momentum and 
$V_{\Delta{\rm E}}$, defined by the difference between the actual orbital 
energy and the orbital energy of a star on the circular guiding center orbit,
is a measure of the eccentricity of the orbit, 
$e=V_{\Delta{\rm E}}/\sqrt{2}V_{\rm LSR}$. The latter assumes a flat Galactic 
rotation curve. $V_{\Delta{\rm E}}$ is also related to the radial action 
integral of the orbit and should be quite robust, even
if the Galactic potential underwent changes in the past.
In order to enhance the overdensities of stars in phase space we have performed 
a three--dimensional wavelet transform with a Mexican hat kernel
${\mathcal{K}}({\bf x}-{\bf x'})=
(3 - ({\bf x}-{\bf x'})^2/a^2) \exp{-(({\bf x}-{\bf  x'})^2/2\,a^2)}$ 
(cf.~Skuljan, Hearnshaw \& Cottrell 1999) 
of the $V_{\rm az} - V_{\Delta{\rm E}} - \nu$ data cube. After some 
experimentation we adopted a wavelet length scale of $a = 20$ km/s which leads 
to the clearest results. The wavelet transform conserves the total
size of the occupied phase space, because the volume integral over 
${\mathcal{K}}$ is equal to zero.

\section{Results and Discussion}
 
\subsection{ Star streams in the halo population}

In Fig.~1 we show scatter plots of slices through the 
three--dimensional phase space distribution of the stars in our sample. 
Each slice is 45$^\circ$ wide, and covers 0$^\circ$ to 45$^\circ$, 
45$^\circ$ to 90$^\circ$, 90$^\circ$ to 135$^\circ$, and 135$^\circ$ to
180$^\circ$, respectively. Corresponding slices through the three-dimensional 
wavelet transforms of the data, which enhance and delineate the overdensities
in a quantitative way, are shown in Fig.~2. A different 
projection of the three--dimensional data cube onto inclination angle $\nu$ 
versus azimuthal velocity $V_{\rm az}$ is shown in Fig.~3. Since the V--shaped 
features in the $V_{\rm az} - V_{\Delta{\rm E}}$ planes are rather narrow, we
have integrated in Fig.~3 all phase space densities over $V_{\Delta{\rm E}}$ . 
A three--dimensional rendering of the data cube is shown in Fig.~4.
As can be 
seen from Figs.~2, 3, and 4 there is a lot of fine structure in the phase space 
distribution of the stars. In the most fully populated part of phase space the 
inclination angles of the orbital planes are about 90 degrees. These stars have 
disk--like kinematics and most of them are even at these comparatively low
metallicities members of the thick disk of the
Galaxy. The diagrams in the upper right and lower left panels of Fig.~2 show 
clearly the Hercules stream at $V_{\rm az}$ = 170 km/s, the stream found by 
Arifyanto \& Fuchs (2006) and Helmi et al.~(2006) at $V_{\rm az}$ = 120 km/s, 
signs of the Arcturus stream at $V_{\rm az}$ = 100 km/s , and possibly a further
feature at $V_{\rm az}$ = 75 km/s, which will be investigated more closely
in a separate study (Klement et al.~, in preparation).

However, our main interest here is star streams in the
genuine halo star population. Tremaine (1993) has estimated a filling factor by 
star streams in the halo around the Sun of about 1, and we expect to 
find a few genuine halo star streams in the sample. In order to isolate these 
we have indicated in Fig.~3 the area populated predominantly by stars with 
disk--like 
kinematics. The dashed lines indicate the loci of stars with $W= \pm 50$ km/s, 
respectively, and $V_{\rm az}$ ranging from 20 to 270 km/s, which should bracket
the velocity range of thick disk stars. Outside this range we find at 
velocities $V_{\rm az} > 0$ three distinct isolated features which we propose 
to interpret as signatures of star streams passing through the solar 
neighbourhood. Two of them (labelled HB and S$_{1}$) are clearly 
visible on the right side of the
bottom right panel of Fig.~2 and in the upper right part of Fig.~3. 
Their centroid positions in phase space are summarized in Table 1. 
Interestingly the feature labelled HB is the stream detected 
by Helmi et al.(1999). Chiba \& Beers (2000) confirmed this discovery 
later using the same data of Beers et al.~(2000) as analyzed in this study, 
but their `clump stars' were isolated in a different way from our method. 
Actually we draw confidence in our search technique from the fact that it 
recovers a previously known star stream in the halo population. The stars in 
the third feature at $V_{\rm az}$ = 140 km/s and $\nu \approx 30^\circ$ have 
the same orbital parameters (cf.~section 3.3) as the stars in the stream newly 
found by Arifyanto \& Fuchs (2006) and Helmi et al.~(2006). The members of
this stream have disk--like kinematics, but comparatively large vertical $W$
velocities. Since the inclination angles $\nu$ depend on the vertical $W$ 
velocity components, such stars will appear as linearly smeared 
features in Figs.~3 and 4. The other linear structures in Figs.~3 and 4 
are, in our view, also due to this effect. The features in the right lower
half in Fig.~3 are therefore very unlikely to be associated with a genuine 
halo stream.
\begin{figure}
\centering
\caption{Projection of the wavelet transform of the phase space distribution of
 halo stars onto a plane spanned by the inclination angles $\nu$ of the
 orbital planes of the stars and their azimuthal velocities $V_{\rm az}$. A 
 similar colour table as in Fig.~2 has been used, but its peak value at red 
 refers to a density which is a factor of about 3 lower than in Fig.~2. The 
 dashed lines indicate the loci of stars with $W=\pm 50$ km/s and $V_{\rm az}$ 
 ranging from 20 to 270 km/s as expected for thick disk stars. The locations of
 stars assigned to halo overdensities are marked by yellow symbols.}
\label{fig3}
\end{figure}
\begin{figure}
\centering
\caption{Three--dimensional rendering of the wavelet transform of the phase 
space distribution of nearby halo stars. The rendered surface corresponds to 
the outer countours in Fig.~2.}
\label{fig4}
\end{figure}

Stars at negative  azimuthal velocities are on retrograde orbits and belong 
exclusively to the halo population. It is more difficult to discern 
overdensities in this part of phase space, because the $V$--shaped branches in
the $V_{\rm az} - V_{\Delta{\rm E}}$ diagrams in Fig.~2 become very narrow and
the  inclinations of the orbital planes are evenly distributed over $\nu$ in 
Fig.~3. However, we find two further isolated features in this part of phase
space. One is located in the upper left frame of
Fig.~2 at negative $V_{\rm az}$ distinctively above the narrow 
$V_{\rm az} - V_{\Delta{\rm E}} $ line of the bulk of the stars. This 
feature is also seen very clearly in the left side of Fig.~3 as a strong 
overdensity at $\nu$ = $6^\circ$. A second feature can be identified on the
left side in the bottom right panel of Fig.~2. Also this feature is seen very
clearly as an overdensity on the left side of Fig.~3 at $\nu$ =
170$^\circ$. The three proposed streams (labelled S$_1$, S$_2$, and S$_3$) are 
described here for the first time, to the best of our knowledge. 
Moreover, we have found signs of Kapteyn's moving group as defined by 
Eggen (1996), although it is not very prominent. Its position in phase space is
given together with that of the other streams in Table 1. Note however that
Kapteyn's star itself is not a member of our sample.
\begin{table}
\caption{Parameters of the proposed halo star streams}
\label{table1}
\centering
\begin{tabular}{crcrrrcccc}
\hline\hline
 & $V_{\rm az}$ & $V_{\Delta{\rm E}}$ & $\nu$ & $N$  & $r_0$ &
     $r_{\rm peri}$ & $r_{\rm apo}$   \\
 & km/s &   km/s &     deg &  & kpc      & kpc   &     kpc  \\
\hline
HB  & 320 &    155  &      150  &   4 & 11      &  7.8  &    20  \\
S$_1$  & 100 &    250  &      165  & 4 & 3.5     &  1.9  &  12 \\
S$_2$  & $-$100 &   520  &      6 &  4 & 3.5     &  1.7  &  17 \\
S$_3$  & $-$100 &   470  &     170 &  10 & 3.7    &   2.1 & 9.1 \\
K   &  $-$75 &   440  &      50   &  3 & 2.8     &  1.5  & 10 \\
\hline
\end{tabular}
\end{table}

\subsection{Monte Carlo simulations}

In Table 1 we give the numbers of stars $N$ which we assign to
each overdensity in phase space by our procedure. These are overplotted as 
yellow symbols in Fig.~3. The numbers of stars in each 
phase space overdensity are quite small. In order to assess the reality of the
overdensities we have performed Monte--Carlo simulations of a smooth velocity
distribution of halo stars. We adopted a triaxial Schwarzschild
distribution with the velocity dispersions determined by Arifyanto \& Fuchs
(2005). The size of the simulated sample was set equal to the observed
number of stars with non--disk kinematics. In a typical Monte - Carlo
simulation the phase space distribution is rather smooth, but does show 
overdensities due to Poisson noise. We have run in
total 100 Monte Carlo simulations and have calculated the residuals of the
wavelet transforms of each individual simulation against the very smooth
superposition of all simulations. For each cell in the 
$V_{\rm az} - V_{\Delta{\rm E}} - \nu$
cube we have determined the mean and variance of the residuals. Finally we have
divided each value of the wavelet transform of the observed phase space
distribution of halo stars shown in Figs.~2, 3, and 4 by the square root of
the variance of the numerically simulated Schwarzschild distribution. This
procedure gives then the S/N ratio of the detected
overdensities. In Fig.~5 a three--dimensional rendering of this ratio at a level
of S/N = 2.8 is shown. At azimuthal velocities less than $V_{\rm az} \leq -$220
km/s the variances of the residuals are nominally so small, that the S/N 
ratio becomes artificially high. Since Fig.~1 shows absolutely no
clustering in this region of phase space, we have masked out this part in 
Fig.~5. Four of the star streams proposed here are clearly discernible, and
are detected at a confidence level of 99.5\% or better. Kapteyn's moving 
group does not show up at this significance level.
\begin{figure}
\centering
\caption{Three--dimensional rendering of the signal - to noise ratio of the 
wavelet transform of the phase space distribution of the halo stars at a level
of S/N = 2.8, which corresponds to a confidence level of 99.5\%. The expected
Poisson noise has been determined with Monte Carlo
simulations of a smooth Schwarzschild distribution (cf.~section 3.2).}
\label{fig5}
\end{figure}

\subsection{Orbital parameters of the halo streams}

We have determined the orbital parameters of the halo streams proposed here 
and list them in Table 1. Since some of the orbits are fairly eccentric, 
Dekker's (1976) approximation of Galactic orbits is not accurate enough to 
give precise estimates of the parameters. This does not, however, diminish its 
power to project even such orbits close together into the same region of phase 
space, if the 
orbits are similar to each other! The mean guiding center radii and inner
and outer turning radii given in Table 1 have been calculated assuming a 
logarithmic Galactic gravitational potential 
$\Phi(r)=V_{LSR}^2\,{\rm ln}(r/1\,{\rm kpc})$ which corresponds to a 
flat rotation curve. The mean guiding center radius of a stream $r_0$ is 
determined by the minimum of the effective potential 
$\Phi_{\rm eff}(r)=\Phi(r)+0.5\cdot L^2/r^2$ and the turning radii are given 
by the condition $E=\Phi_{\rm eff}(r_{\rm peri},r_{\rm apo})$, where $E$ 
denotes the orbital energy. As can be seen from Table 1 stream HB is at present
close to its perigalacticon, whereas the other streams dive deep into the inner 
Galaxy. It would be intriguing if distant counterparts of the locally detected 
streams could be identified. For instance, Mart\'inez-Delgado et al.~(2007)
have presented a numerical simulation of the Sagittarius stream which suggests 
that its leading arm passes in the vicinity of the Sun out of the direction of 
the North Galactic Pole nearly perpendicular through the Galactic plane,
although this was questioned by Newberg et al.~(2007). 
The stream labelled S$_3$, which we propose here, has precisely these 
kinematics. However, its orbital parameters deduced here are difficult to 
reconcile with the simulated morphology of the Sagittarius stream. Obviously 
this depends to a large degree on the details of the model adopted for the 
Galactic potential. Observations as presented here might be 
helpful to constrain these.
\begin{acknowledgements}
We are grateful to the anonymous referee for his helpful comments. BF and CF 
thank the Mt.~Stromlo observatory for its hospitality, where this research was 
begun. CF acknowledges financial support by the Academy of Finland.
\end{acknowledgements}

\end{document}